\documentclass[twocolumn,showpacs,preprintnumbers,amsmath,amssymb,floatfix]{revtex4}
\usepackage{graphicx}
\usepackage{epsfig}
\usepackage{amsfonts}
\usepackage{amssymb}
\usepackage{epsf}
\newcommand{\insertplot}[5]{\begin{figure}
 \hfill\hbox to 0.05in{\vbox to #5in{\vfill
 \inputplot{#1}{#4}{#5}}\hfill}
 \hfill\vspace{-.1in}
 \caption{#2}\label{#3}
 \end{figure}}
\newcommand{\inputplot}[3]{
 \special{ps: plotfile #1}
}

\newcounter{fig}   \newcommand{\lbfig}[1]{\refstepcounter{fig}
\label{#1} }

\newcommand{\vphi}{\varphi}

\begin{document}

\title{
Wormholes in Dilatonic Einstein-Gauss-Bonnet Theory}

\author{ 
{\bf Panagiota Kanti}
}
\affiliation{ Division of Theoretical Physics, Department of Physics,\\
University of Ioannina, Ioannina GR-45110, Greece}

\author{
{\bf Burkhard Kleihaus, Jutta Kunz}
}
\affiliation{
{Institut f\"ur Physik, Universit\"at Oldenburg,
D-26111 Oldenburg, Germany}
}
\date{\today}
\pacs{04.70.-s, 04.70.Bw, 04.50.-h}
\begin{abstract}
We construct traversable wormholes in dilatonic Einstein-Gauss-Bonnet
theory in four spacetime dimensions,
without needing any form of exotic matter.
We determine their domain of existence,
and show that these wormholes satisfy a generalized Smarr relation.
We demonstrate linear stability with respect to radial perturbations
for a subset of these wormholes.
\end{abstract}
\maketitle

\noindent{\textbf{~~~Introduction.--~}}
When the first wormhole, the  ``Einstein-Rosen bridge",
was discovered in 1935 \cite{Einstein-Rosen} 
as a feature of Schwarzschild geometry, 
it was considered a mere mathematical curiosity of the theory.
In the 1950s, Wheeler showed \cite{Wheeler} that a 
wormhole can connect not only two different universes 
but also two distant regions of our own Universe. 
However, the dream of interstellar travel shortcuts was shattered
by the following findings: 
(i) the Schwarzschild wormhole is dynamic - its ``throat" 
expands to a maximum radius and then contracts again to zero circumference
so quickly that not even a particle moving at the
speed of light can pass through \cite{Kruskal}, 
(ii) the past horizon of the Schwarz\-schild geometry 
is unstable against small perturbations -
the mere approaching of a traveler would change it to a
proper, and thus impenetrable, one \cite{stability}.

However, in 1988 Morris and Thorne \cite{Morris-Thorne} 
found a new class of wormhole solutions 
which possess no horizon, and thus could be traversable.
The throat of these wormholes is kept open by a type of matter
whose energy-momentum tensor violates the energy conditions. 
A phantom field, a scalar field with a
reversed sign in front of its kinetic term, was shown to be a suitable
candidate for the exotic type of matter necessary to support traversable wormholes
\cite{phantom}.

In order to circumvent the use of exotic matter
to obtain traversable wormholes,
one is led to consider generalized theories of gravity.
Higher-curvature theories of gravity are 
suitable candidates to allow for the existence of
stable traversable wormholes.
In particular, the low-energy heterotic string effective theory 
\cite{Gross:1986mw,Metsaev:1987zx}
has provided the framework for such a generalized gravitational theory 
in four dimensions where the curvature term $R$ of Einstein's theory 
is supplemented by the presence of additional fields
as well as higher-curvature gravitational terms. 
The dilatonic Einstein-Gauss-Bonnet (DEGB) theory 
offers a simple version that contains, 
in addition to $R$, a quadratic curvature term, the Gauss-Bonnet (GB) term,
and a scalar field (the dilaton) coupling exponentially to the GB term,
so that the latter has a nontrivial contribution to the
four-dimensional field equations.

Here we investigate the existence of wormhole solutions
in the context of the DEGB theory. 
No phantom scalar fields or other exotic forms of matter 
are introduced. 
Instead, we rely solely on the existence of the higher-curvature GB term 
that follows naturally from the compactification of the 
ten-dimensional heterotic superstring theory down to four dimensions. 

\noindent{\textbf{~~~DEGB theory.--~}}
We consider the following effective action 
\cite{Mignemi:1992nt,Kanti:1995vq,Chen:2009rv,Kleihaus:2011tg}
motivated by the low-energy heterotic string theory
\cite{Gross:1986mw,Metsaev:1987zx}
\begin{eqnarray}  
S=\frac{1}{16 \pi}\int d^4x \sqrt{-g} \left[R - \frac{1}{2}
 \partial_\mu \phi \,\partial^\mu \phi
 + \alpha  e^{-\gamma \phi} R^2_{\rm GB}   \right],
\label{act}
\end{eqnarray} 
where 
$\phi$ is the dilaton field
with coupling constant $\gamma$, $\alpha $ is a  positive numerical
coefficient given in terms of the Regge slope parameter,
and
$R^2_{\rm GB} = R_{\mu\nu\rho\sigma} R^{\mu\nu\rho\sigma}
- 4 R_{\mu\nu} R^{\mu\nu} + R^2$ 
is the GB correction. 

Here we consider only static, spherically-symmetric
solutions of the field equations. 
Hence we may write the spacetime line element
in the form \cite{Kanti:1995vq}
\begin{equation}
ds^2 
= -e^{\Gamma(r)}dt^2+e^{\Lambda(r)}dr^2
+r^2\left(d\theta^2+\sin^2\theta d\varphi^2 \right).
\label{metricS} 
\end{equation}

In \cite{Kanti:1995vq} it was demonstrated that DEGB theory 
admits static black hole solutions, based on this line element.
But it was also observed that, besides the black hole
solutions, the theory admits other classes of solutions. 
One of the examples presented showed a pathological behavior for
the $g_{rr}$ metric component and the dilaton field at a finite
radius $r=r_0$ but had no proper horizon with $g_{tt}$ being regular
for all $r \geq r_0$.  
Since the solution did not exhibit any singular behavior 
of the curvature invariants at $r_0$, 
it was concluded that the pathological behavior was due to the 
choice of the coordinate system.

Here we argue that this class of asymptotically flat solutions is indeed
regular and represents a class of wormholes 
with $r_0$ being the radius of the throat.
Indeed, the coordinate transformation $r^2  =l^2 + r_0^2$ 
leads to a metric without any pathology,
\begin{equation}
ds^2 =  -e^{2\nu(l)}dt^2+f(l)dl^2
+(l^2+r_0^2)\left(d\theta^2+\sin^2\theta d\varphi^2 \right) \ .
\label{metricL} 
\end{equation}
In terms of the new coordinate, the expansion at the throat $l=0$, yields 
$f(l)  =  f_0 + f_1 l + \cdots$,
$e^{2\nu(l)}  =  e^{2\nu_0}(1 +\nu_1 l)  + \cdots$,
$\phi(l)      =  \phi_0 + \phi_1 l +  \cdots$,
where $f_i$, $\nu_i$ and $\phi_i$ are constant coefficients. 
All curvature invariants, including the GB term,
remain finite for $l\to 0$.

The expansion coefficients $f_0$, $\nu_0$ and $\phi_0$ are free parameters,
as well as  
the radius of the throat $r_0$ and the value of $\alpha$ --
the value of the constant $\gamma$ is set to 1 in the calculations. 
The set of equations remains invariant under the simultaneous changes 
$\phi \rightarrow \phi +\phi_*$ and $(r,l) \rightarrow (r,l) e^{-\phi_*/2}$. 
The same holds for the changes 
$\alpha \rightarrow k \alpha$ and $\phi \rightarrow \phi + \ln k$. 
As a result, out of the parameter set $(\alpha, r_0, \phi_0)$ 
only one is independent: we thus fix the value of $\phi_0$, 
in order to have a zero value of the dilaton field at infinity, 
and create a dimensionless parameter $\alpha/r_0^2$ out of the remaining two. 
Also, since only the derivatives of the metric function $\nu$ 
appear in the equations of motion, we fix the value of $\nu_0$ 
to ensure asymptotic flatness at radial infinity. Note that
force-free wormhole solutions, i.e., with $\nu(l)\equiv 0$, cannot exist with
a nonphantom scalar, as was shown in \cite{Bronnikov:2009az}.

Of particular interest is the constraint on the value of the first derivative
of the dilaton field at the throat, which originates from the 
diagonalization of the dilaton and Einstein equations in the 
limit $l \to 0$. In terms of the expansions it
translates into a constraint on the value of the parameter $\phi_1$, i.e.
\begin{equation}
\phi_1^2 = \frac{f_0(f_0-1)}{2\alpha\gamma^2 e^{-\gamma \phi_0}
\left[f_0-2(f_0-1)\frac{\alpha}{r_0^2} e^{-\gamma \phi_0}\right]}\ .
\label{rel-l0}
\end{equation}
Since the left-hand-side of the above equation is positive-definite, we must
impose the constraint $f_0 \geq 1$. This constraint introduces a boundary
in the phase space of the wormhole solutions.
The expression inside the square brackets 
in the denominator remains positive and has no roots if 
$a/r_0^2<e^{\phi_0}/2$ -
this inequality is automatically satisfied for the set of solutions presented.

For $l \to \infty$ we demand asymptotic flatness for the two metric
functions and a vanishing dilaton field. Then, the corresponding 
asymptotic expansion yields
$\nu  \rightarrow  -\frac{M}{l} + \cdots $,
$f    \rightarrow  1+ \frac{2M}{l} + \cdots $,
$\phi  \rightarrow  -\frac{D}{l} + \cdots $
where $M$ and $D$ are identified with the mass and dilaton charge of the
wormhole, respectively. Unlike the case of the black hole solutions \cite{Kanti:1995vq},
the parameters $M$ and $D$ characterizing the wormholes
at radial infinity are not related, in agreement with the classification of
this group of solutions as two-parameter solutions. 

\noindent{\textbf{~~~Wormhole properties.--~}}
A general property of a wormhole is the existence of a throat,
i.~e., a surface of minimal area (or radius for spherically symmetric 
spacetimes). Indeed, this property is implied by the form
of the line element (\ref{metricL}) above, with $f(0)$ and $\nu(0)$ 
finite.
To cast this condition in a coordinate independent way, we define the 
proper distance from the throat by 
$
\xi = \int_0^l \sqrt{g_{_{ll}}} dl' = \int_0^l \sqrt{f(l')} dl'$.
Then the conditions for a minimal radius 
$
\left.\frac{dr}{d\xi}\right|_{l=0} = 0$,
$ \left.\frac{d^2r}{d\xi^2}\right|_{l=0} > 0$
follow from the substitution of the expansion at the throat.

In order to examine the geometry of the space manifold, 
we consider the isometric embedding
of a plane passing through the wormhole. 
Choosing the $\theta=\pi/2$ plane, we set
$
f(l) dl^2 +(l^2+r_0^2) d\vphi^2 = dz^2 + d\eta^2 +\eta^2 d\vphi^2,
$
where $\{z,\eta,\varphi\}$ are a set of cylindrical coordinates in the three-dimensional
Euclidean space $R^3$. Regarding $z$ and $\eta$ as functions of $l$, we find 
$\eta(l)$ and $z(l)$.
We note that the curvature radius of the curve $\{\eta(l),z(l)\}$ at $l=0$ is 
given by $R_0 = r_0 f_0$. From this equation we obtain an  
independent meaning for the parameter $f_0$ as the ratio 
of the  curvature radius and the radius of the throat, $f_0 = R_0/r_0$.

Essential for the existence of the wormhole solution 
is the violation of the null energy condition
$
T_{\mu\nu} n^\mu n^\nu \geq $0, 
for any null vector field $n^\mu$.
For spherically symmetric solutions, this condition can be expressed as
$
-G_0^0+G_l^l \geq  0$ and  $-G_0^0+G_\theta^\theta\geq  0$, 
where the Einstein equations have been employed. 
The null energy condition is violated in some region 
if one of these conditions does not hold. 
By using the expansion of the fields near the throat, we find that there 
\begin{equation}
\left[-G_0^0+G_l^l\right]_{l=0} = -\frac{2}{f_0 r_0^2} < 0 \ , 
\end{equation}
provided $e^{2\nu(0)}\neq 0$, i.e. in the absence of a horizon.

The wormhole solutions satisfy a Smarr-like mass formula 
\begin{eqnarray}
M  & =  & 2 S_{\rm th} \frac{\kappa}{2\pi} -\frac{D}{2\gamma} \nonumber\\
   &    &
     +\frac{1}{8 \pi \gamma}\int{\sqrt{-g}g^{ll}\frac{d\phi}{dl}
                    \left(1+2\alpha \gamma^2 e^{-\gamma\phi} \tilde{R}\right)}d^2x 
\nonumber
\end{eqnarray}
where $\kappa$ denotes the surface gravity at the throat and
$$ 
S_{\rm th} = \frac{1}{4}\int{ \sqrt{h}\left(1+2\alpha e^{-\gamma\phi} 
 \tilde{R}\right) d^2x }\ . 
$$
Here $h_{\mu\nu}$ is the induced metric on the throat, $\tilde{R}$ the scalar
curvature of $h$, and the integral is evaluated at $l=0$.
Thus the known DEGB mass formula for black holes \cite{Kleihaus:2011tg}
is augmented by a contribution which may be interpreted 
as a modified throat dilaton charge, where the GB modification
is of the same type as the GB modification of the area (or entropy, in case of
black holes).

\noindent{\textbf{~~~Numerical results.--~}}
For the numerical calculations we use the line element (\ref{metricL}).
At $l=0$ regularity requires relation (\ref{rel-l0}) to hold
with $f(0) = f_0$ or $\phi(0)=\phi_0$ in order to obtain a unique solution.
Note that $f \to 1$ for $l\to \infty$ is always satisfied. 
Thus the asymptotic boundary conditions read
$\nu \to 0 \ {\rm and} \ \phi \to 0$.

The field equations, reducing to a system of ODE's, were solved for $1.0001 \leq f_0 \leq 20.0$  
fixing $\gamma = 1$. Wormhole solutions were found for every value of $\alpha/r_0^2$
considered up to 0.13 - this translates into a lower bound on the radius of the throat $r_0$
that can be arbitrarily large. In Fig.~1, we show
the metric and dilaton functions for an indicative set of wormhole solutions.
In Fig.~2,  we present the isometric embedding of the solution with $\alpha/r_0^2=0.02$ and
 $f_0=1.1$.

\begin{figure}[t]
\lbfig{f-1}
\begin{center}
\includegraphics[height=.23\textheight, angle =0]{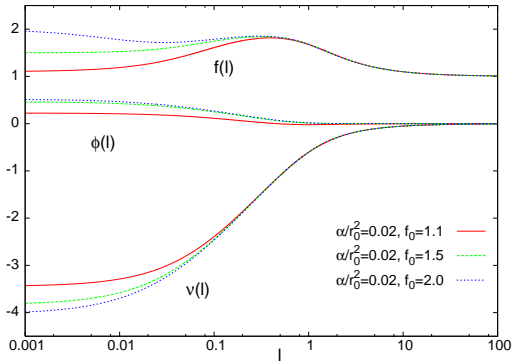}
\end{center}
\caption{
The metric and dilaton functions for
$f_0=1.1$ (red or solid), $1.5$ (green or dashed ), $2.0$ (blue or dotted), 
and $\alpha/r_0^2=0.02$, versus $l$.}
\end{figure}

\begin{figure}[h!]
\lbfig{f-2}
\centerline{
\includegraphics[height=.28\textheight, angle =0]{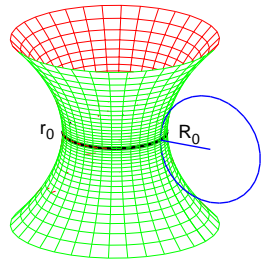}
\hspace{1cm}}
\vspace{-0.8cm}
\caption{
Isometric embedding for a wormhole solution
for $\alpha/r_0^2=0.02$ and $f_0=1.1$.
}
\end{figure}

In Fig.~3
we exhibit the domain of existence of the wormhole solutions.
Here we show the scaled area of the throat $A/16 \pi M^2$ 
versus the scaled dilaton charge $D/M$ for several values of $\alpha/r_0^2$.
We observe that the domain of existence is bounded by
three curves indicated by dots, crosses and asterisks. 
The boundary indicated by asterisks corresponds to the limit $f_0=1$ and
coincides with the black hole curve  since $-g_{00}(r_0)$ tends to zero 
in this limit.
The boundary indicated by crosses corresponds to the limit $f_0 \to \infty$. In this
limit, the mass $M$ and the dilaton charge $D$ assume finite values.
The same holds for the redshift function, the dilaton field and all curvature
invariants at the throat.
The third boundary, indicated by dots, is characterized by curvature singularities;
it  emerges when branches of solutions with fixed $\alpha/r_0^2$ 
terminate at singular configurations with the 
derivatives of the functions developing a discontinuity at some point $l_{\rm crit}$ 
outside the throat. 

\begin{figure}[b]
\lbfig{f-3}
\begin{center}
\includegraphics[height=.27\textheight, angle =0]{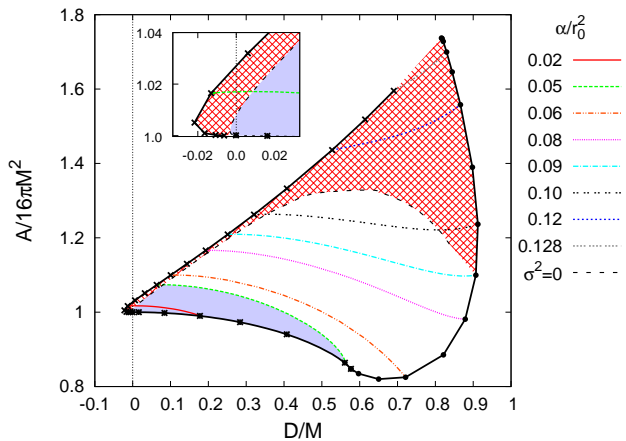}
\end{center}
\caption{
The scaled area of the throat versus
the scaled dilaton charge for several values of $\alpha/r_0^2$. 
The domain of existence is bounded by the black hole solutions
(asterisks), the $f_0 \to \infty$ solutions (crosses),
and curvature singularity solutions (dots).
The shaded areas indicate linear stability (lilac or lower), instability (red or upper),
undecided yet (white) with respect to radial perturbations.}
\end{figure}

\noindent{\textbf{~~~Stability.--~}}
A crucial requirement for traversable wormhole solutions is their stability.
We therefore assess the stability of the DEGB wormhole solutions
with respect to radial perturbations.
We allow the metric and dilaton functions to depend on both $l$ and $t$, and 
we decompose them into an unperturbed part and the perturbations:
$ \tilde{\nu}(l,t)  =  \nu(l) +\lambda \delta\nu(l) e^{i\sigma t} $,
$ \tilde{f}(l,t)  =  f(l) +\lambda \delta f(l) e^{i\sigma t} $,
$ \tilde{\phi}(l,t)  =  \phi(l) +\lambda \delta\phi(l) e^{i\sigma t} $,
where 
$\lambda$ is considered as small.
Substituting the above in the (time-dependent) Einstein and dilaton equations and
linearizing in $\lambda$, we obtain a system of linear ODEs for the functions 
$\delta\nu(l)$, $\delta f(l)$ and $\delta\phi(l)$.

Rearranging appropriately the system of ODEs, we derive a decoupled second-order
equation for $\delta\phi$,  
\begin{equation}
(\delta \phi)'' + q_1 (\delta\phi)'+ (q_0+q_\sigma \sigma^2)\, \delta\phi = 0 \ , 
\label{lineq1}
\end{equation}
where $q_1$, $q_0$ and $q_\sigma$ depend on the unperturbed solution. 
All coefficients take up constant values as $l \to \infty$ and, in order to ensure
normalizability, $\delta\phi$ is demanded to vanish in that limit. However, while
$q_\sigma$ is bounded at $l=0$, $q_1$ and $q_0$ 
diverge as $1/l$. 
To avoid this singularity, we consider the transformation 
$\delta \phi = F(l) \psi(l)$, where $F(l)$ satisfies 
$F'/F = -q_1(l)/2$. This yields 
\begin{equation}
\psi'' + Q_0\psi +\sigma^2 q_\sigma \psi = 0 \, 
\label{lineq2}
\end{equation}
where $Q_0=-q'_1/2-q_1^2/4+q_0$ is bounded at $l=0$.
\begin{figure}[t]
\lbfig{f-4}
\begin{center}
\includegraphics[height=.23\textheight, angle =0]{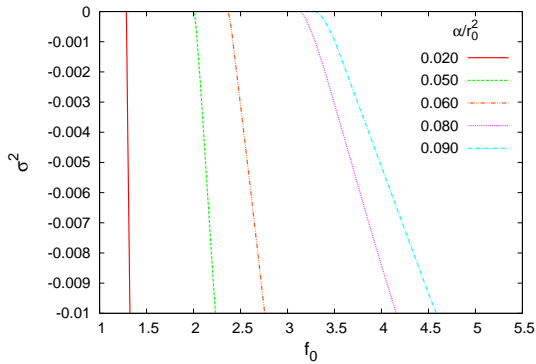}
\end{center}
\caption{
The eigenvalue $\sigma^2$ versus $f_0$ for several values of 
$\alpha/r_0^2$.}
\end{figure}
Then, for regular, normalizable solutions,
$\psi$ has to vanish both asymptotically and at $l=0$ .

We solved the ODEs 
together with the normalization constraint for several values
of $\alpha/r_0^2$ and $f_0$. 
A solution exists only for certain values of the eigenvalue $\sigma^2$.
In Fig.~4, we exhibit the negative modes of
families of wormhole solutions, which are thus unstable.
However, there is also a region in parameter space, where no negative
mode exists, i.e. a region where the
DEGB wormhole solutions are linearly stable with respect to radial perturbations.
The different solutions are also marked in the domain
of existence (Fig.~3).

\noindent{\textbf{~~~Junction conditions.--~}}
When we extend the wormhole solutions to the second asymptotically
flat part of the manifold ($l \rightarrow -\infty$) in a symmetric way, jumps
appear in the derivatives of the metric and dilaton functions at $l=0$. These
can be attributed to the presence of matter
at the throat of the wormhole. Introducing the action 
$\int (\lambda_1 + \lambda_0 2 \alpha e^{-\phi_0} \bar{R})\,\sqrt{-\bar{h}}\,d^3x$,
where $\bar{h}_{ab}$ is the induced metric on the throat and $\bar{R}$ the corresponding Ricci scalar, 
the junction conditions take the form
\begin{eqnarray}
\hspace*{-0.5cm}
\rho-\lambda_0\,\frac{4 \alpha e^{-\phi_0}}{r_0^2}-\lambda_1&=&
\frac{8 \alpha e^{-\phi_0}}{r_0^2}\,\frac{\phi_0'}{\sqrt{f_0}}\,, \\
p+\lambda_1 &=& \frac{2\nu'_0}{\sqrt{f_0}}\,,\\
\hspace*{-0.5cm} \lambda_0\,\frac{4 \alpha e^{-\phi_0}}{r_0^2} +\frac{\rho_{\rm dil}}{2}
&=& 
\left(\phi'_0 +\frac{8 \alpha e^{-\phi_0}}{r_0^2}\,{\nu_0'}\right)\frac{1}{\sqrt{f_0}}\,.
\end{eqnarray}
We have also 
assumed that the matter at the throat takes the form of a perfect
fluid and a dilaton charge density $\rho_{\rm dil}$. A simple numerical analysis
shows that the junction conditions can be easily satisfied for normal matter
with positive energy density and pressure, 
for appropriately chosen constants $\lambda_1$ and $\lambda_0$. 
The presence of $\rho_{\rm dil}$ ensures that the stability considerations are not
affected by the matter on the throat. 

\noindent{\textbf{~~~Conclusions.--~}}
Traversable wormholes do not exist in general relativity, unless some
exotic matter is introduced. However, if string theory corrections are
taken into account the situation changes dramatically. 
Here we investigated wormhole solutions in DEGB theory without introducing
exotic matter. We determined the domain of existence and showed linear stability
with respect to radial perturbations for a subset of solutions. 
Since the radius of the throat is bounded from below only, 
the wormholes can be arbitrarily large. Astrophysical consequences will
be addressed in a forthcoming paper as well as the existence of stationary
rotating wormhole solutions in the DEGB theory. 

\noindent{\textbf{~~~Acknowledgements.--~}}
We gratefully acknowledge discussions with Eugen Radu.
B.K.~acknowledges support by the DFG.


\begin{thebibliography}{99}


\bibitem{Einstein-Rosen}
  A.~Einstein and N.~Rosen,
  Phys.\ Rev.\  {\bf 48} (1935) 73.

\bibitem{Wheeler} J.A. Wheeler, {\it Geometrodynamics} (Academic, New York, 1962).

\bibitem{Kruskal}
  M.~D.~Kruskal,
  Phys.\ Rev.\  {\bf 119}, 1743 (1960);
  R.~W.~Fuller and J.~A.~Wheeler,
  Phys.\ Rev.\  {\bf 128}, 919 (1962).

\bibitem{stability} I.~H.~Redmount, Prog. Theor. Phys. {\bf 73} (1985) 1401;
D.~M.~Eardley, Phys. Rev. Lett. {\bf 33} (1974) 442; R.~M.~Wald and
S. Ramaswamy, Phys. Rev. {\bf D21} (1980) 2736.


\bibitem{Morris-Thorne} M.~S.~Morris and K.~S.~Thorne,
  Am.\ J.\ Phys.\  {\bf 56}, 395 (1988).

\bibitem{phantom} H.~G.~Ellis,
  J.\ Math.\ Phys.\  {\bf 14}, 104 (1973);
  K.~A.~Bronnikov,
  Acta Phys.\ Polon.\  B {\bf 4}, 251 (1973);
  T.~Kodama,
  Phys.\ Rev.\  D {\bf 18}, 3529 (1978);
  C.~Armendariz-Picon,
  Phys.\ Rev.\  D {\bf 65}, 104010 (2002).


\bibitem{Gross:1986mw}
  D.~J.~Gross, J.~H.~Sloan,
  Nucl.\ Phys.\  {\bf B291 } (1987)  41.

\bibitem{Metsaev:1987zx}
  R.~R.~Metsaev, A.~A.~Tseytlin,
  Nucl.\ Phys.\  {\bf B293 } (1987)  385.
  
\bibitem{Kanti:1995vq}
  P.~Kanti, N.~E.~Mavromatos, J.~Rizos, K.~Tamvakis and E.~Winstanley,
  Phys.\ Rev.\  D {\bf 54} (1996) 5049;  Phys.\ Rev.\  D {\bf 57} (1998) 6255.

\bibitem{Mignemi:1992nt}
  S.~Mignemi and N.~R.~Stewart,
  Phys.\ Rev.\  D {\bf 47} (1993) 5259.

\bibitem{Chen:2009rv}
  C.~M.~Chen, D.~V.~Gal'tsov, N.~Ohta, D.~G.~Orlov,
  Phys.\ Rev.\  {\bf D81 } (2010)  024002.

\bibitem{Kleihaus:2011tg}
  B.~Kleihaus, J.~Kunz, E.~Radu,
  Phys.\ Rev.\ Lett.\  {\bf 106}, 151104 (2011).

\bibitem{Bronnikov:2009az}
  K.~A.~Bronnikov, E.~Elizalde,
  Phys.\ Rev.\  {\bf D81 } (2010)  044032.


\end{thebibliography}
\end{document}